\documentclass[twocolumn,showpacs,preprintnumbers,amsmath,amssymb]{revtex4}

\usepackage{graphicx}
\usepackage{dcolumn}
\usepackage{bm}

\begin{document}


\title{Effects of anharmonic vibration on large-angle
  quasi-elastic scattering of $ ^{16}$O+$^{144}$Sm}
\author{Muhammad Zamrun F.}
\email{zamrun@nucl.phys.tohoku.ac.jp}
\author{K. Hagino}
\email{hagino@nucl.phys.tohoku.ac.jp}
\affiliation{Department of Physics, Tohoku University,
Sendai 980-8578, Japan}

\date{\today}

\begin{abstract}
We study the effects of double octupole and quadrupole phonon
excitations in the $^{144}$Sm nucleus on quasi-elastic 
$^{16}$O$+^{144}$Sm scattering at 
backward angles. 
To this end, we use the coupled-channels
framework, taking into account explicitly 
the anharmonicities of the
vibrations. We use the same coupling scheme as that previously 
employed to explain 
the experimental data of sub-barrier fusion cross sections for the
same system. We show that the experimental data for the quasi-elastic
cross sections are well reproduced in this way, although the
quasi-elastic barrier distribution has a distinct high energy
peak which is somewhat smeared in the experimental barrier 
distribution. We also discuss the
effects of proton transfer on the quasi-elastic 
barrier distribution. 
Our study indicates that the fusion and quasi-elastic barrier
distributions for this system cannot be accounted for simultaneously
with the standard coupled-channels approach. 
\end{abstract}


\maketitle

\section{Introduction}

The effect of channel coupling, that is, 
couplings of the relative motion
between the colliding nuclei to their intrinsic motions as well as 
transfer reactions, have been well known 
in heavy-ion collisions around the Coulomb barrier. 
In heavy-ion fusion reactions at sub-barrier energies, the channel 
coupling effects enhance considerably the fusion cross sections as compared 
to the prediction of potential model calculation 
\cite{beck-88,bah-tak98,das-98}. 
It has been well established by now that the
channel coupling gives rise to a distribution of
potential barriers \cite{esb-81,nag-86,hag-95}. Based on this idea, a
method was proposed to extract barrier distributions directly from
experimental fusion excitation functions by taking the second derivative of the
product of center mass energy, $E$, and the fusion cross section,
$\sigma_{\rm fus}(E)$, with respect to $E$ \cite{row-91}.
Coupled-channels calculations as well as high precision fusion data 
have shown that the fusion barrier
distributions, $D^{\rm fus}=d^2[E\sigma_{\rm fus}(E)]/dE^2$, is sensitive to
the details of channel couplings, while the sensitivity is much more
difficult to see in the fusion cross sections \cite{das-98,das-981,leigh-95}.

Similar information as the fusion cross section can also be
obtained 
from the quasi-elastic scattering (a sum of elastic, inelastic and
transfer processes) at backward angles \cite{ARN88}. 
Timmers {\it et al.} 
measured the quasi-elastic scattering cross section for several
systems \cite{thim-95}, 
for which the fusion barrier distribution had already 
been extracted \cite{leigh-95}. 
They proposed that the corresponding 
barrier distribution can be extracted 
by taking the first derivative of the ratio of the quasi-elastic
 to the Rutherford cross sections, $d\sigma_{\rm qel}/d\sigma_R$, with
respect to the energy, $E$, {\it i.e.,} 
$D^{\rm qel}=-d(d\sigma_{\rm qel}/d\sigma_R)/dE$. 
The properties of the quasi-elastic barrier distributions have been studied
in more details in Ref. \cite{hag-04}. These studies show that 
the quasi-elastic barrier distribution is similar to the 
fusion barrier distribution, although the former is somewhat smeared 
and less sensitive to the nuclear structure effects. 
\begin{figure}
\includegraphics{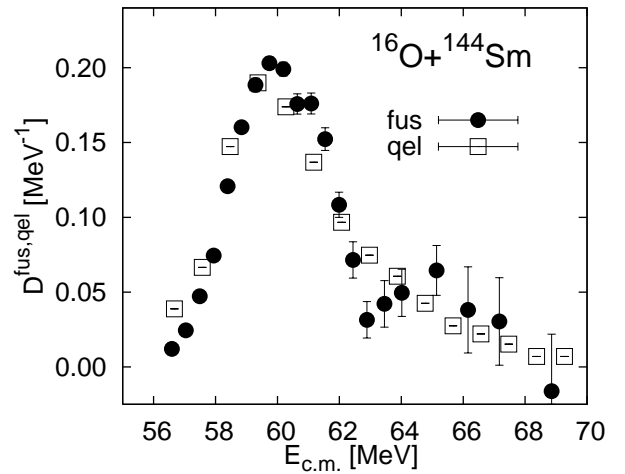}
\caption{Comparison between the experimental fusion (the filled circles) 
and quasi-elastic (the open squares) barrier
  distributions for the  $^{16}$O$+^{144}$Sm reaction. They are
  normalized to unit area in the energy interval between 
$E_{\rm c.m.}=$ 56 and 70 MeV. 
The experimental data are
  taken from Refs. \cite{leigh-95} and \cite{thim-95}.}  
\end{figure}

One of the systems which Timmers {\it et al.} measured is 
$^{16}$O$+^{144}$Sm \cite{thim-95}. 
Figure 1 shows the comparison of the  
experimental barrier distribution extracted from fusion (the filled
circles) and quasi-elastic (the open squares) processes. 
In order to compare the two barrier distributions, 
we scale them so that the energy integral between $E_{\rm c.m.}$= 56 
and 70 MeV is unity. For energies below 62 MeV, 
the two barrier distributions resemble each other. 
However, at higher energies, 
they behave rather differently, although 
the overall width of the distributions is similar to each other. 
That is, the quasi-elastic barrier distribution 
decreases monotonically as a function of energy while 
the fusion barrier distribution exhibits a distinct peak at energy around
$E_{\rm c.m.}=65$ MeV. 
So far, no theoretical calculations have succeeded in explaining 
this difference. 
The coupled-channels calculations of Timmers {\it et al.} \cite{thim-95}
with the computer code {\tt ECIS} \cite{ecis}, which took 
into account the one 
quadrupole, $2^+$, and the one octupole, $3^-$, 
phonon excitations of $^{144}$Sm, 
were unable to reproduce both the
experimental data of the quasi-elastic cross sections and 
the quasi-elastic barrier distribution.
The {\tt ECIS} results for the ratio of quasi-elastic scattering 
to the Rutherford cross sections fall off more steeply than  
the experimental data, while the obtained barrier distribution 
has a secondary peak similar
to the fusion barrier distribution.  
They argued that this failure is largely due to the residual 
excitations not 
included in the {\tt ECIS} calculations, which they 
postulated to be transfer channels. 
Esbensen and Buck have also performed the coupled-channels
calculations for this system taking into account the second order couplings
\cite{Esb-96}. However, they did not analyze the quasi-elastic 
barrier distribution. 

These previous coupled-channels calculations took into account 
only the single phonon excitations in $^{144}$Sm. 
On the other hand,   
Hagino {\it et al.} \cite{hag-97,hag-971} have shown that 
the double anharmonic quadrupole and octupole 
phonon excitations play an important role in reproducing the experimental 
fusion barrier distribution for this system. 
However, its effect on the quasi-elastic
scattering 
has not yet been clarified so far. The aim of this paper is then to  
study weather the double anharmonic vibrational excitations of the 
$^{144}$Sm nucleus can explain the difference in the shape of barrier 
distribution between fusion and quasi-elastic. 
The role of proton 
transfer reactions in this system is also discussed.

The paper is organized as follows. In the next section, we briefly
explain the coupled-channels formalism which takes 
into account the anharmonicities of the 
vibrational excitations. We present the results of our calculations 
in Sec. III. We then summarize the paper in Sec. IV.

\section{Coupled-channels formalism for anharmonic vibration}

In this section, we briefly describe the coupled-channels formalism
which includes the effects of anharmonic excitations of the
vibrational states.  
We follow the procedure of Refs. \cite{hag-97,hag-971}, which was 
successfully applied to describe the experimental fusion cross
sections as well  
as the fusion barrier distributions of $^{16}$O+$^{144,148}$Sm systems. 
The total Hamiltonian of the system is assumed to be
\begin{eqnarray}
H&=&-\frac{\hbar^2}{2\mu}\nabla^2+H_{\rm vib}+V_{\rm coup}(\boldsymbol{r},\xi)
\end{eqnarray}
where $\boldsymbol{r}$ is the coordinate of the relative motion between the 
target and the projectile nuclei, $\mu$ is the reduced mass and $\xi$
represents the internal vibrational degrees of freedom of the target
nucleus. $H_{\rm vib}$ describes the 
vibrational spectra in the target nucleus.

The coupling between the relative motion and the intrinsic motion of
the target nucleus is described by the coupling potential $V_{\rm coup}$ 
in Eq.(1), which consists 
of the Coulomb and nuclear parts. Using the no-Coriolis 
(iso-centrifugal) approximation 
\cite{bah-tak98,hag-99}, they are given as
\begin{eqnarray}
V_{\rm coup}(r,\xi)=V_C(r,\xi)+V_N(r,\xi),\qquad\qquad\qquad\qquad \\
V_C(r,\xi)=\frac{Z_PZ_Te^2}{r}\left(1+\frac{3R_T^2}{5r^2}
\frac{\hat{O}_{20}}{\sqrt{4\pi}}+\frac{3R_T^3}{7r^3}
\frac{\hat{O}_{30}}{\sqrt{4\pi}}\right), 
\label{vcoupc}
\\
V_{N}(r,\xi)=\frac{-V_0}{\left[1+\textrm{exp}\left(\frac{
[r-R_0-R_T(\hat{O}_{20}+\hat{O}_{30})/\sqrt{4\pi}]}{a}\right)\right]}.
\quad\,\label{vcoupn}
\end{eqnarray}\\
Here $\hat{O}_{20}$ and $\hat{O}_{30}$ are the excitation operators for
the quadrupole and octupole vibrations, respectively, and $R_T$ is 
the target radius. 
The effect of
anharmonicities for the quadrupole and octupole vibrations are taken into
account based on the U(5) limit of the 
Interacting Boson Model (IBM). The
matrix elements of the operator 
$\hat{O}=\hat{O}_{20}+\hat{O}_{30}$ 
in Eqs.(\ref{vcoupc}) and (\ref{vcoupn}) then read 
\cite{baha-9394,hag-97,hag-971},
\begin{widetext} 
\begin{equation}
O_{ij}=
\left[\begin{array}{cccccc}
0&\beta_2&\beta_3&0&0&0\\
\beta_2&-\frac{2}{\sqrt{14N}}\chi_2 \beta_2&-\frac{2}{\sqrt{15N}}\chi_3\beta_3&
\sqrt{2(1-1/N)}\beta_2&\sqrt{1-1/N}\beta_3&0\\
\beta_3&-\frac{2}{\sqrt{15N}}\chi_3\beta_3&
-\frac{2}{\sqrt{21N}}\chi_{2f}\beta_2&0&
\sqrt{1-1/N}\beta_2&
\sqrt{2(1-1/N)}\beta_3\\
0&\sqrt{2(1-1/N)}\beta_2&0&-\frac{4}{\sqrt{14N}}\chi_2\beta_2&
-\sqrt{\frac{8}{15N}}\chi_3\beta_3&0\\ 
0&\sqrt{1-1/N}\beta_3&\sqrt{1-1/N}\beta_3&-\sqrt{\frac{8}{15N}}\chi_3\beta_3&
(-\frac{2}{\sqrt{14N}}\chi_2-\frac{2}{\sqrt{21N}}\chi_{2f})\beta_2&
-\sqrt{\frac{8}{15N}}\chi_3\beta_3\\
0&0&\sqrt{2(1-1/N)}\beta_3&0&-\sqrt{\frac{8}{15N}}\chi_3\beta_3&
-\frac{4}{\sqrt{21N}}\chi_{2f}\beta_2\\
\end{array} \right]
\end{equation}
\end{widetext}
for 6 low-lying states ($i,j$=1-6), where 
$|1\rangle = |0^+\rangle$, 
$|2\rangle = |2^+\rangle$, 
$|3\rangle = |3^-\rangle$, 
$|4\rangle = |2^+\otimes2^+\rangle$, 
$|5\rangle = |2^+\otimes3^-\rangle$, and
$|6\rangle = |3^-\otimes3^-\rangle$. 
In Eq.(5), $\beta_2$ and $\beta_3$ are
the quadrupole and the octupole deformation parameters, respectively, 
which can be estimated from the electric transition probabilities. 
The scaling of
coupling strength with $\sqrt{N}$, $N$ being the number of boson in
the system, is introduced to ensure the
equivalence between the IBM and the geometric
model in the large $N$ limit \cite{baha-9394}. 
When all the $\chi$ parameters in Eq.(5)  
are set to be zero then the quadrupole moment of all the states
vanishes, and one obtains the harmonics 
limit in the large $N$ limit. Nonzero values of $\chi$ generate the
quadrupole moments, and, together with finite boson number, they are
responsible for the anharmonicities in the vibrational excitations. 

\section{$^{16}$O$+^{144}$Sm reaction : Comparison with experimental data}

We now apply the formalism to analyze the quasi-elastic
scattering data of $^{16}$O$+^{144}$Sm \cite{thim-95}. The
calculations are performed with a version \cite{hag2} of the coupled-channels
code {\tt CCFULL} \cite{hag-99} once the coupling matrix elements 
are determined from Eq.(5).
Notice that the iso-centrifugal approximation employed in this code 
works well for
quasi-elastic scattering at backward angles \cite{hag-04}. In the code,
the regular boundary condition is imposed at the origin instead of the
incoming wave boundary condition. 

\subsection{Effect of anharmonicities of nuclear vibrations}

In the calculations presented below, we include only the excitations 
in the $^{144}$Sm nucleus whilst the excitations of the $^{16}$O 
is not explicitly included. For sub-barrier fusion reactions, 
the latter has been shown to lead only to a shift of the fusion
barrier distribution in energy 
without significantly altering its shape \cite{hag-972}, and 
hence can be incorporated in the choice of the bare potential. 
This is a general feature for reactions with the $^{16}$O as a
projectile. We have confirmed that it is 
the case also for the quasi-elastic barrier distribution. That is, 
although the $^{16}$O excitations contribute to the 
absolute value of quasi-elastic 
cross sections themselves, the shape of quasi-elastic barrier
distribution is not altered much. Since we are interested 
mainly in the difference of the shape between the fusion and the 
quasi-elastic barrier distributions, we simply 
do not include the $^{16}$O excitations and instead adjust the 
inter-nuclear potential. 

For simplicity, we take the eigenvalues of the $H_{\rm vib}$ in Eq.(1)
to be $\epsilon=n_2\epsilon_2+n_3\epsilon_3$, 
where $n_2$ and $n_3$ are the number of quadrupole and octupole
phonons, respectively. $\epsilon_2$ and $\epsilon_3$ are 
the excitation energies of the quadrupole
and the octupole phonon states of the target nucleus,
{\it i.e.}, $\epsilon_2=1.61$ MeV and $\epsilon_3=1.81$ MeV, respectively. 
Notice that we assume the harmonic spectra for the phonon
excitations. It has been shown in Refs. \cite{hag-97,hag-971} 
that the effect of anharmonicity with respect to the excitation energy 
on the barrier distribution is insignificant once the energy of the 
single phonon states is fixed. The radius and diffuseness parameters
of the real part of the nuclear potential are taken to be 
the same as those in Ref. \cite{hag-97}, 
{\it i.e.,} $r_{0}=1.1$ fm and 
$a=0.75$ fm, respectively, while the depth parameter 
is slightly adjusted in order to reproduce 
the experimental quasi-elastic cross sections.  
The optimum value is obtained as $V_0=112$ MeV.  
As usually done, we use a short-range imaginary potential 
with $W_{0}=30$ MeV, $r_{w}=1.0$ fm and $a_w=0.3$ fm to simulate the
compound nucleus formation. Finally, the target radius is taken to be
$R_T=1.06A_T^{1/3}$.  We use the same values for the parameters 
$\beta_2,\beta_3, N, \chi_2, \chi_{2f}$, and $\chi_3$ as in 
Ref. \cite{hag-97}. All the calculations presented below are 
performed at $\theta_{\rm c.m.}=170^\circ$.   
\begin{figure}
\includegraphics{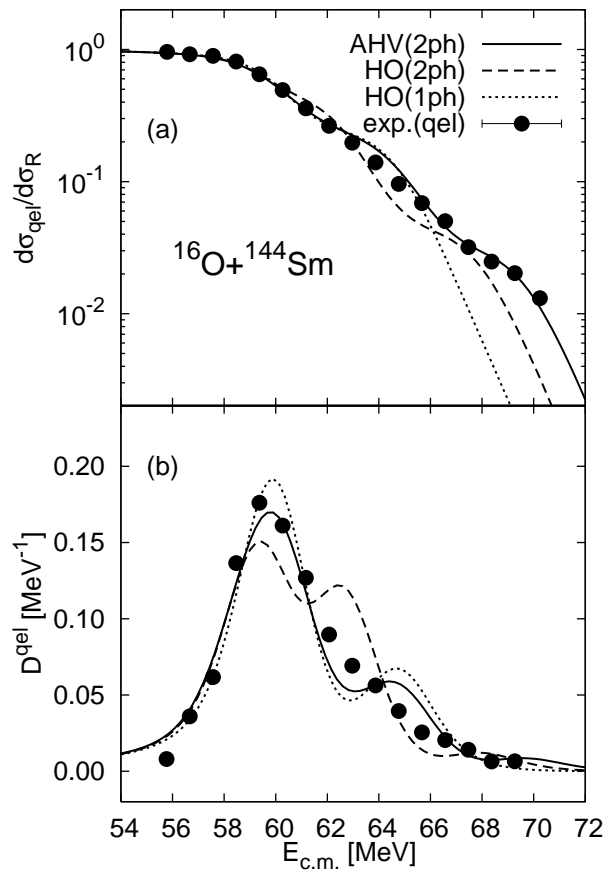}
\caption{Comparison of the experimental data 
  with the coupled-channels calculations for  $^{16}$O$+^{144}$Sm
  reaction for (a) the ratio of quasi-elastic  
to the Rutherford cross sections and for (b) quasi-elastic barrier
distribution. The dotted and dashed lines are obtained by including 
up to the single and the double phonon excitations in the harmonic 
limit, respectively. The solid line is the result of the 
coupled-channels calculations with the double anharmonic phonon excitations. 
The experimental data are taken from Ref. \cite{thim-95}.}
\end{figure}
\begin{figure}
\includegraphics{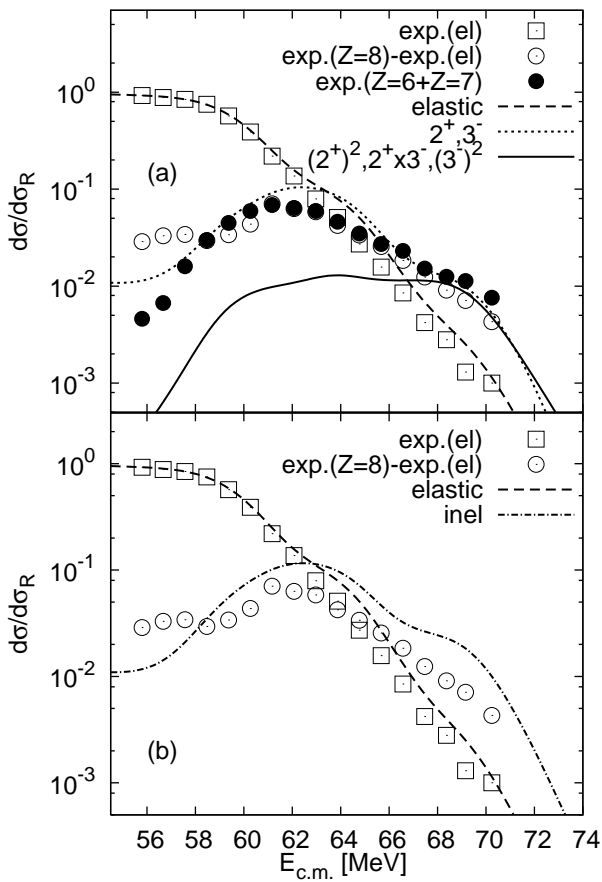}
\caption{(a) 
Comparison of the measured pure elastic (the open squares), the 
$Z=8\,(-\,\textrm{el})$ (the open circles) and the residual (the
filled circles) components of $d\sigma_{\rm qel}/d\sigma_R$ with the 
coupled-channels calculations for $^{16}$O$+^{144}$Sm reaction. 
The $Z=8\,(-\,\textrm{el})$ component is defined as the $Z=8$ yields
subtracted the elastic component, while the residual component the 
sum of $Z=6$ and 7 yields. 
The dashed line is the result of elastic scattering, while the dotted 
line shows the inelastic cross sections for the single 2$^+$ and 3$^-$
phonon states. The solid line is the result of the sum of inelastic 
cross sections for the double phonon states in $^{144}$Sm. 
(b) The same as (a) 
  but for the pure elastic and the total inelastic cross sections. 
The experimental data are
  taken from Ref. \cite{thim-95}.} 
\end{figure}

The results of the coupled-channels calculations are compared with the
experimental data in Fig. 2. Figures 2(a) and 2(b) show the ratio of the
quasi-elastic to the Rutherford cross sections,
$d\sigma_{\rm qel}/d\sigma_R$, and the quasi-elastic barrier
distributions,  $D^{\rm qel}$, respectively. The dotted line denotes the
result in the harmonic limit, 
where coupling to the quadrupole and octupole vibrations in
$^{144}$Sm are truncated at the single phonon level, {\it i.e.,} only the
$2^+$ and $3^-$ states are taken into account and all the
$\chi$ parameters in Eq.(5) are set to be zero. 
As we see this calculation fails to reproduce the
experimental data. The obtained  quasi-elastic cross sections, 
$d\sigma_{\rm qel}/d\sigma_R$, drop much faster than the experimental 
data at high energies. Also the quasi-elastic barrier distribution, 
$D^{\rm qel}$, exhibits a distinct peak at energy around 
$E_{\rm c.m.}=65$ MeV. These results are similar to the one achieved
in Ref. \cite{thim-95}. The dashed line represents the result when the
coupling to the quadrupole and octupole vibrations of $^{144}$Sm is
truncated at the double phonon states in the harmonic limit. In this
case, we take into account the couplings to the $2^+$, $3^-$,
$2^+\otimes2^+$,$2^+\otimes 3^-$ and \mbox{$3^-\otimes3^-$} states. 
It is obvious that the results are inconsistent with the  experimental data. 
To see the effect of anharmonicities of the vibrations, we then perform the
the same calculations using the coupling matrix elements given in Eq.(5). 
The resultant quasi-elastic excitation
function and the quasi-elastic barrier distribution are shown 
by the solid line. 
The calculated ratio of 
quasi-elastic to Rutherford cross sections quite well agree with the 
experimental data. 
This suggests that the inclusion of anharmonic effects in the 
vibrational motions is important for the description of the 
quasi-elastic excitation functions for the $^{16}$O$+^{144}$Sm reaction. 
On the other hand, 
the result for $D^{\rm qel}$ is still similar to the 
barrier distribution obtained by assuming the harmonic limit 
truncated at the one 
phonon level (the dotted line), although 
the former has a more smooth peak. 

Figure 3 shows  the decomposition of the quasi-elastic 
cross sections to each channel for the calculation with 
the coupling to the double anharmonic vibrations 
(the solid line in Fig. 2).  
The fraction of cross section for each channel $i$ in the
quasi-elastic cross section, 
$d\sigma_i/d\sigma_{\rm qel}=d\sigma_i/[\sum_jd\sigma_j]$, 
is also shown in Fig. 4. The open squares are the experimental elastic cross
section while the open circles are the measured excitation function for 
$Z=8$ subtracted the contribution from the elastic channel.
The latter contains not only
the neutron transfer components but also the contributions of 
inelastic cross sections. The filled circles 
are the experimental residual (a sum of $Z=7$ and
$Z=6$ yields) components of the
$d\sigma_{\rm qel}/d\sigma_R$. The dashed line shows results of the
coupled-channels 
calculations for the elastic channel.  It reproduces reasonably well the
experimental data for elastic scattering. The $Z=8$ component of 
quasi-elastic cross sections is almost exhausted by the single phonon
excitations, that is, the combined $2^+$ and $3^-$ channels, as 
shown by the dotted-line in Figs. 3(a) and 4(a). 
The cross sections for the double phonon channels are given by the
solid line in Figs. 3(a) and 4(a).  
These are important at energies higher than around 66 MeV. 
If the components of all the inelastic 
channels included in the calculation are summed up, 
we obtain the dot-dashed line in Figs. 3 (b) and 4(b). 
\begin{figure}
\includegraphics{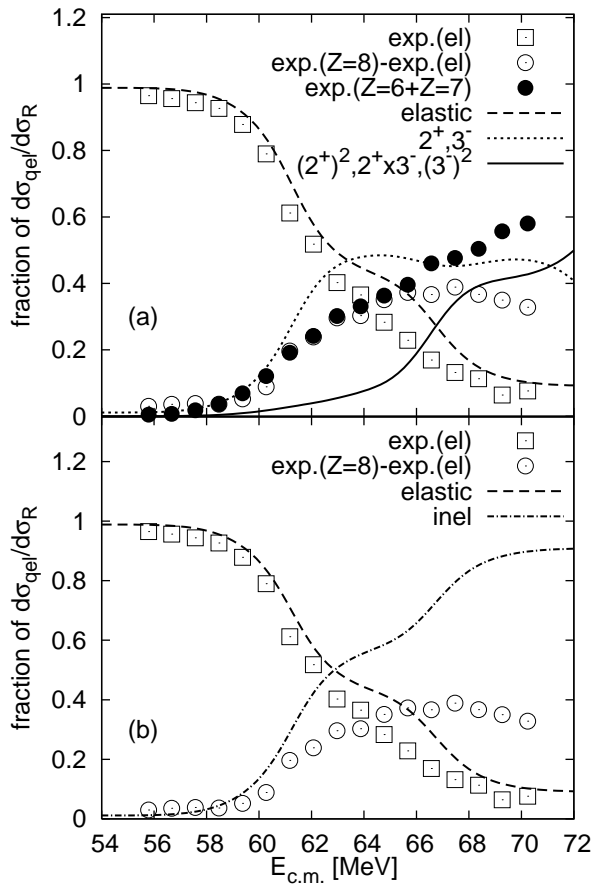}
\caption{ Same as Fig. 3, but for the fraction of cross section for
  each channel  in the quasi-elastic cross sections. }
\end{figure}

\subsection{Effects of proton transfer reactions}

In the previous subsection we have shown that the experimental 
quasi-elastic cross sections can be well
explained within the present coupled-channels calculations, which 
takes into account only the inelastic excitations in $^{144}$Sm. 
However, the shape of quasi-elastic barrier
distribution is still somewhat inconsistent with the experimental data. 
As one sees in Figs. 3(a) and 4(b), the experimental data indicate that 
the charged particle transfer reactions 
may also play some role (see the filled circles in the figures). 
In this subsection, we therefore investigate the effects of 
proton transfer reactions, in addition to the 
anharmonic double phonon excitations. 
To this end, we use the macroscopic 
form factor for the transfer coupling \cite{dasso-8586}, 
\begin{equation}
F_{\rm trans}(r)=-F_{\rm tr}\frac{dV(r)}{dr}
\end{equation}
where $F_{\rm tr}$ is the coupling strength and $V(r)$ is the real part
of the nuclear potential.
In this paper, we consider a single proton transfer as well as 
the direct proton pair transfer reactions, although the experimental 
Z=6 component may also include the alpha-particle transfer channel. 
The corresponding optimum
$Q-$values for the transfer between the ground states 
are $Q_{\rm opt}(1p)=-1.79$ MeV and $Q_{\rm opt}(2p)=0.13$ MeV,
respectively. 
The coupling strength $F_{\rm tr}$ in Eq.(6) 
is determined so that the experimental transfer cross sections 
for each Z=6 and Z=7 components \cite{thim-thes} are reproduced. 
The optimum values for $F_{\rm tr}$ are found to be  0.12 and 0.16 fm for
the one and the two proton transfer channels, respectively. 
\begin{figure}
\includegraphics{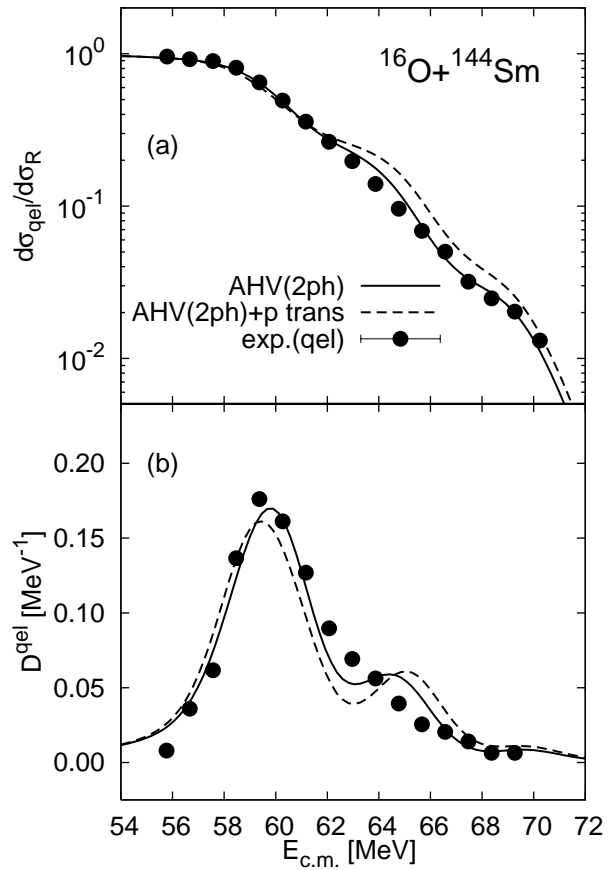}
\caption{Effect of proton transfers on the quasi-elastic scattering 
cross sections (the upper panel) and on the quasi-elastic barrier 
distribution (the lower panel) for $^{16}$O$+^{144}$Sm reaction. 
The solid line is the result of the coupled-channels calculations 
including the effect of double anharmonic vibrations only. 
The dashed line is obtained by including, in addition, 
the couplings to the proton transfer channels. 
The experimental data are taken from Ref. \cite{thim-95}.}
\end{figure}
\begin{figure}
\includegraphics{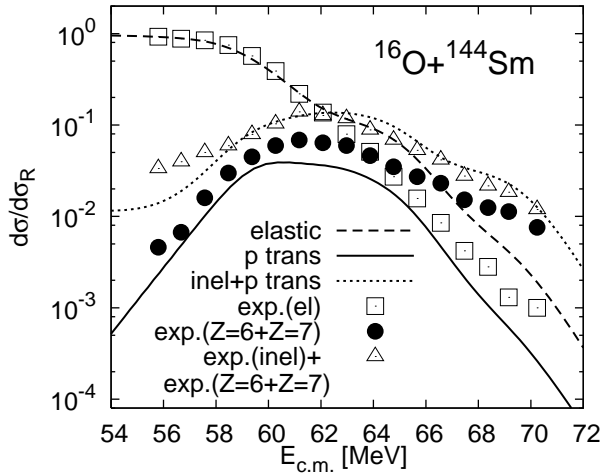}
\caption{
Contribution of quasi-elastic cross sections from several channels. 
The solid and dashed line are the results of the 
coupled-channels calculations for 
the proton transfer and the elastic cross sections, respectively. 
The dotted line denotes the sum of total inelastic 
and proton transfer cross sections. 
The corresponding experimental data are shown by the filled circles, 
the open squares, and the open triangles, respectively, which are 
taken from Ref. \cite{thim-95}.}
\end{figure}
\begin{figure}
\includegraphics{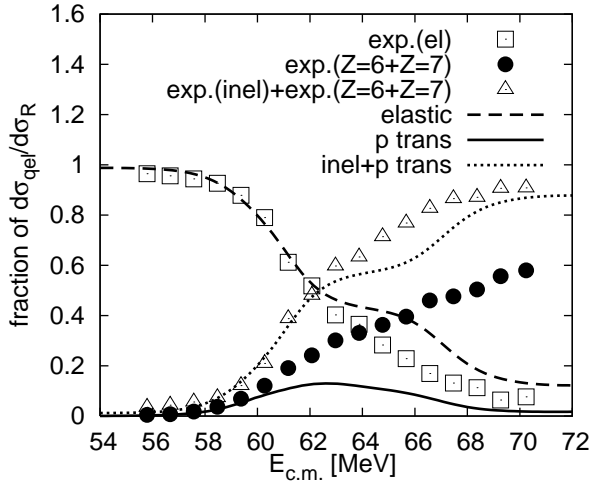}
\caption{
Same as Fig. 6, but for the fraction in the quasi-elastic 
cross sections. } 
\end{figure}
 \begin{figure}
\includegraphics[angle=0,width=0.457\textwidth]{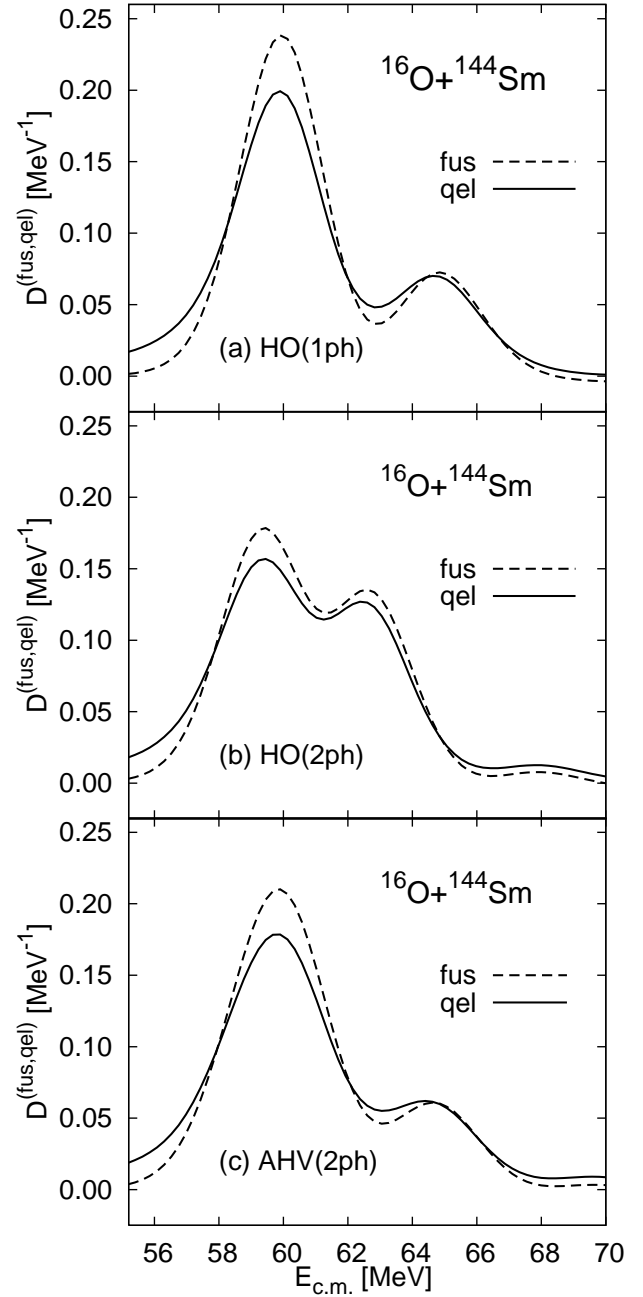}
\caption{Comparison of the theoretical fusion barrier distribution 
(dashed line) with the  quasi-elastic barrier distribution (solid-line) 
obtained with different coupling schemes for
  $^{16}$O$+^{144}$Sm system. Both functions are normalized to unit
  area in energy interval between 54 and 70 MeV. (a) The results of  
the coupling to one phonon state of quadrupole and 
octupole excitations of $^{144}$Sm in the harmonic oscillator limit.
(b) The same as (a) but for the coupling up to double phonon states. 
(c) The result when the coupling to anharmonic vibration of double
quadrupole and octupole excitations in $^{144}$Sm is taken into account.}
\end{figure}

The effects of proton transfer reactions on the quasi-elastic scattering 
is illustrated in Fig. 5. The solid line represents the results of the 
calculations including only the coupling to the double 
anharmonic vibrations. The dashed line is obtained by taking the
coupling to the proton transfer channels into account, in
addition to the anharmonic vibration channels. 
The upper panel shows the quasi-elastic cross sections, while the 
lower panel the quasi-elastic barrier distribution. 
We observe from Fig. 5(a) that the inclusion of proton transfer reactions 
overestimates the experimental $d\sigma_{\rm qel}/d\sigma_R$ at
energies between 62 and 68 MeV. Also the 
higher peak in the quasi-elastic
barrier distribution becomes more distinct and thus worsens as
compared to the calculation without the transfer channels. 

Figure 6 shows the contribution of each channel to the quasi-elastic 
cross sections. The fraction of each contribution is also shown in 
Fig. 7. The open squares are the
experimental elastic cross sections, while the filled circles and the open
triangles are the experimental proton transfer cross sections and the
sum of total inelastic and transfer cross sections, respectively. 
The coupled-channels calculations for the elastic cross sections are
shown by the dashed-line. 
Although it reproduces the experimental data below around 62 MeV, it 
overestimates the data at higher energies. 
The sum of the contributions from the
total inelastic and the proton transfer channels is denoted by the 
dotted line, which reproduces the experimental data reasonably well, 
although the proton transfer cross sections themselves are 
underestimated  at energies larger than 60 MeV (the solid line). 
The overestimation of the quasi-elastic cross section indicated in
Fig. 5(a) is therefore largely due to the contribution of elastic channel. 

From this study, we conclude that the inclusion of the 
proton transfer reactions in the coupled-channels calculations 
does not explain the difference of the shape between the fusion 
and quasi-elastic barrier distributions for the 
$^{16}$O$+^{144}$Sm system.

\subsection{Discussions}

We have argued that the presence of high energy shoulder, instead
of high energy peak, in the quasi-elastic barrier distribution for 
the scattering between 
$^{16}$O and $^{144}$Sm nuclei cannot be accounted for within the present 
coupled-channels calculations, which take into account the anharmonic 
double phonon excitations in $^{144}$Sm as well as the proton transfer 
channels. 
Figure 8 compares the calculated 
fusion barrier distribution $D^{\rm fus}$ and the
 corresponding quasi-elastic barrier distribution $D^{\rm qel}$ 
for several coupling schemes 
as shown in Fig. 2 in the coupled-channels calculations. 
The solid line shows the quasi-elastic barrier 
distribution while the dashed line is for the fusion barrier
distribution. They are normalized so that the
energy integral between 54 and 70 MeV is unity. 
Figures 8(a) and 8(b) are obtained by including the one phonon and the
two phonon  excitations in $^{144}$Sm in the 
harmonic limit, respectively. Figure 8(c) is the result of 
the double anharmonic vibration coupling. 
From these figures, it is evident that the theoretical fusion and
quasi-elastic barrier distributions 
are always similar to each other within the same coupling scheme, 
although the latter is slightly more smeared due to the low-energy
tail \cite{hag-04}. 
This would be the case even with the excitations in $^{16}$O as well 
as neutron transfer channels, which are not included in the present 
coupled-channels calculations. 
Therefore, it seems unlikely that the experimental fusion and
quasi-elastic barrier distributions can be explained simultaneously 
within the standard coupled-channels approach. 

\section{Conclusion}

We have studied the effects of double anharmonic vibrations 
of the $^{144}$Sm nucleus on the large angle quasi-elastic scattering for
$^{16}$O$+^{144}$Sm system. We have 
shown that the experimental data for the quasi-elastic scattering
cross sections for this reaction can be reasonably well explained.  
However, we found that the obtained quasi-elastic barrier
distribution still shows the clear doubled-peaked structure, 
that is not seen in the experimental data. 
This was not resolved even if we took the proton transfer channels 
into account. Our coupled-channels calculations indicate 
that, within the same coupling scheme, the quasi-elastic and fusion barrier 
distributions are always similar to each other. 
Although detailed analyses including neutron transfer channels in 
a consistent manner are still necessary, it is thus unlikely that 
the fusion and quasi-elastic barrier distributions can be explained 
simultaneously with the standard coupled-channels framework. 
This fact might be related to the large diffuseness problem in sub-barrier 
fusion, in which dynamical effects such as couplings to deep-inelastic 
scattering are one of the promising origins
\cite{NBD04,DHN04,MHD07}. 
It is still an open problem to perform the coupled-channels
calculations with such dynamical effects and explain the difference of
the shape between the fusion and the quasi-elastic barrier distributions 
for the $^{16}$O$+^{144}$Sm reaction. 

\begin{acknowledgments}
This work was partly supported by The 21st Century Center of
Excellence Program ``Exploring New Science by Bridging
Particle-Matter Hierarchy'' of Tohoku University 
and partly by Monbukagakusho Scholarship 
and Grant-in-Aid for Scientific Research under
the program number 19740115 from the Japanese Ministry of
Education, Culture, Sports, Science and Technology.
\end{acknowledgments}

\end{document}